%% file: 00-spec-hpc2021.tex
\pdfoutput=1

\documentclass[conference]{IEEEtran}
\IEEEoverridecommandlockouts

\usepackage{placeins} 
\usepackage[font=small,skip=2pt]{caption}

\usepackage{cite}
\usepackage{lipsum,graphicx,multicol}
\usepackage{amsmath,amssymb,amsfonts}
\usepackage{algorithmic}
\usepackage{graphicx}
\usepackage{textcomp}
\usepackage{xcolor}
\usepackage{comment}
\usepackage{color}
\usepackage{cite}
\usepackage{capt-of}
\usepackage{subfigure}
\def\BibTeX{{\rm B\kern-.05em{\sc i\kern-.025em b}\kern-.08em
    T\kern-.1667em\lower.7ex\hbox{E}\kern-.125emX}}
 
\usepackage{threeparttable}
\usepackage[colorinlistoftodos]{todonotes}
\usepackage{hyperref}
\usepackage{multirow}
\hypersetup{
    colorlinks=true,
    linkcolor=blue,
    filecolor=magenta,      
    urlcolor=cyan,
}
\usepackage{url}
\newcommand{\fc}[1]{\color{magenta}#1} 

\newcommand\spec[1]{SPEC}
\newcommand\spechpg[1]{SPEC HPG}
\newcommand\spechpc[1]{SPEC HPC2021}
\newcommand\hpcbeta[1]{HPC2021 Beta}

\begin{document}
\bstctlcite{IEEEexample:BSTcontrol}

\pagenumbering{roman}

\title{First Experiences in Performance Benchmarking with the New SPEChpc 2021 Suites}

\author{\IEEEauthorblockN{Holger Brunst\IEEEauthorrefmark{3}, Sunita Chandrasekaran\IEEEauthorrefmark{1}, Florina Ciorba\IEEEauthorrefmark{2}, Nick Hagerty\IEEEauthorrefmark{6}, Robert Henschel\IEEEauthorrefmark{8}, Guido Juckeland\IEEEauthorrefmark{5}, \\ Junjie Li\IEEEauthorrefmark{7}, Ver\'onica G. Melesse Vergara\IEEEauthorrefmark{6}, Sandra Wienke\IEEEauthorrefmark{4}, and Miguel Zavala\IEEEauthorrefmark{1},
\thanks{Authors listed in alphabetical order. Following~\cite{credit} to list contributions: \textbf{H.Brunst} (Investigation, Writing - Review \& Editing),
\textbf{S. Chandrasekaran} (Conceptualization, Investigation, Supervision, Project administration, Writing - Original Draft)
\textbf{F. Ciorba} (Methodology, Writing - Review \& Editing),
\textbf{N. Hagerty} (Software, Writing - Review \& Editing)
\textbf{R. Henschel} (Writing - Review \& Editing),
\textbf{G. Juckeland} (Data Curation, Writing - Review \& Editing)
\textbf{J. Li} (Software, Investigation, Writing - Original Draft)
\textbf{V. Vergara} (Software, Investigation, Writing - Review \& Editing)
\textbf{S. Wienke} (Investigation, Writing - Original Draft)
\textbf{M. Zavala} (Software)}}
\IEEEauthorblockA{\IEEEauthorrefmark{1}
University of Delaware, Newark, DE, USA. 
\IEEEauthorrefmark{2}University of Basel, Basel, Switzerland}
\IEEEauthorblockA{\IEEEauthorrefmark{3}Technische Universit\"at Dresden, Dresden, Germany. 
\IEEEauthorrefmark{4}RWTH Aachen University, Aachen, Germany}
\IEEEauthorblockA{\IEEEauthorrefmark{5}Helmholtz-Zentrum Dresden-Rossendorf, Dresden, Germany.  
\IEEEauthorrefmark{6}Oak Ridge National Laboratory, Oak Ridge, TN, USA}
\IEEEauthorblockA{\IEEEauthorrefmark{7} Texas Advanced Computing Center, The University of Texas at Austin, Austin, TX, USA. }
\IEEEauthorrefmark{8}{Indiana University, Bloomington, IN, USA}
}

\maketitle
\thispagestyle{plain}
\pagestyle{plain}
\pagenumbering{arabic}


\input{abstract}


\section{Introduction}\label{sec:Intro}
\input{01-intro}

\section{Motivation}\label{sec:Motivation}
\input{02-motivation}

\section{Overview of SPEChpc 2021 benchmark suites}\label{sec:Overview}
\input{03-overview-benchmark}

\section{SPEC harness}\label{sec:Harness}
\input{04-harness}

\section{Results}\label{sec:Results}
\input{05-results}


\section{Related Work}\label{sec:Rel-work}
\input{07-related}

\section{Conclusions and Future Work}\label{sec:Conclusions}
\input{08-conclusion-future-work}

\section{Acknowledgements}

The authors would like to acknowledge a number of facilities and grants for their support, but first and foremost, we would like to thank the full SPEC HPG group for the tremendous effort behind developing and releasing the SPEChpc 2021 benchmark suites. Then, the Center for Information Services and HPC at TU Dresden for providing its facilities for high throughput calculations; RWTH Aachen University under project rwth0663 for supporting simulations on their computing resources; supported by NSF under grant no. 1814609; Gauss Centre for Supercomputing e.V. for funding this project by providing computing time through the John von Neumann Institute for Computing (NIC) on the GCS Supercomputer JUWELS at Jülich Supercomputing Centre (JSC); Oak Ridge Leadership Computing Facility, a DOE Office of Science User Facility supported under Contract DE-AC05-00OR22725 for resources used; the Frontera supercomputer at TACC funded by NSF for large scaling calculations and profiling; the Swiss PASC initiative via the SPH-EXA project; the Swiss State Secretariat for Education, Research and Innovation (SERI).

\bibliographystyle{IEEEtran}
{\footnotesize \bibliography{ref}}

\end{document}

%% file: abstract.tex
\begin{abstract}

Modern HPC systems are built with innovative system architectures and novel programming models to further push the speed limit of computing. 
The increased complexity poses challenges for performance portability and performance evaluation. 
The Standard Performance Evaluation Corporation (SPEC) has a long history of producing industry-standard benchmarks for modern computer systems. 
SPEC’s newly released SPEChpc~2021 benchmark suites, developed by the High Performance Group, are a bold attempt to provide a fair and objective benchmarking tool designed for state-of-the-art HPC systems.  
With the support of multiple host and accelerator programming models, the suites are portable across both homogeneous and heterogeneous architectures. 
Different workloads are developed to fit system sizes ranging from a few compute nodes to a few hundred compute nodes.  
In this manuscript, we take a first glance at these benchmark suites and evaluate their portability and basic performance characteristics on various popular and emerging HPC architectures, including x86 CPU, NVIDIA GPU, and AMD GPU. 
This study provides a first-hand experience of executing the SPEChpc~2021 suites at scale on production HPC systems, discusses real-world use cases, and serves as an initial guideline for using the benchmark suites.
\end{abstract}

%% file: 01-intro.tex

Evaluating the performance of computing systems using carefully designed benchmarks supports comparisons between different systems. 
Performance benchmarks have contributed to improvements in successive generations of systems, which are important for pushing the speed limit of computing, 
purchasing investments,
development of research software and performance analysis tools, 
system maintenance, and others. 
%

The SPEC High Performance Group (HPG)~\cite{spechpg} has been designing benchmarks for the last three decades, releasing the first benchmark suite, SPEC HPC96, in 1996. 
Over the years, SPEC HPG released various benchmark suites that target all parallel execution layers of modern HPC systems.
Nevertheless, each suite focused on individual parallelism layers: SPEC~MPI2007 on inter-node communication, SPEC~OMP2012 on intra-node CPU parallelism, and SPEC~ACCEL 2017 on the performance of accelerator devices.

\textbf{Motivation.} 
The design and release of SPEC~ACCEL raised the question as to \textit{how to measure performance of a system with multiple accelerator devices}.
To answer this question, contributors of SPEC~HPG discussed how to better reflect \textit{\textbf{overall~system~performance}}, specifically considering the increasing heterogeneity in system architectures and diversity in programming models.
The outcome materialized as the launch of a new application search program~\cite{specsearchprogram} in late 2017 to gather input from the HPC community on potential benchmark applications that are, among others, characterized by \textit{more than one form of parallelism} (cross-node, node-level, with/out accelerator offloading). 
The SPEChpc~2021 benchmark suites are comprised of those applications that fulfill the selection criteria (see Section~\ref{sec:Overview}).

Existing benchmarks either focus on low-level hardware performance features and provide microbenchmarks and small application kernels, or on higher level code features and provide mini- or proxy apps selected to prepare the hardware and software stacks of upcoming large systems.
In contrast, SPEChpc~2021 comprises real-world applications solicited from the broader HPC community, provides a set of execution and reporting rules, and adopts a peer-review process before publishing benchmarking results online. 

To the best of our knowledge, SPEChpc~2021 is a \textit{one of a kind benchmark suite} that offers a harness to handle the process from installation to ensuring the correctness of the results and providing a performance score (called the SPEC score) to enable ranking, that explores hybrid programming models as well as MPI-only, and facilitates benchmarking on university clusters and large HPC center systems.

\textbf{Contributions.} This work brings forward the following contributions:
(1)~Presents a detailed overview of the new SPEChpc~2021 benchmark suites, including code statistics, instruction mix, MPI call percentages, and roofline models.
%
(2)~Provides a first study of how well the suites meet the search requirements by evaluating the performance results on a variety of hardware systems as well as exploring different system configurations. 
To the best of our knowledge, this work is among the first few, with respect to evaluating scientific applications on a pre-exascale system, namely Spock, equipped with AMD MI100 GPUs.
(3)~Compares the performance of employing MPI+X programming models with various X, namely OpenMP host, OpenACC and OpenMP target offloading. 

\textbf{Impact.}
The significance and impact of this performance benchmarking study are multi-fold: 
(i)~Added value through extensive testing on a wide range of platforms, going beyond classical building and compilation or error testing.
(ii)~The performance numbers obtained across devices allow identifying unsuspected system configuration bugs.
(iii)~Engaged the HPC community by enlisting the help of testers for the beta release candidate of the suites.

\textbf{Paper organization.} 
The motivation behind performance benchmarking with the new suites is presented in Section~\ref{sec:Motivation}.
Section~\ref{sec:Overview} provides an overview of the new SPEChpc~2021 suites, including applications, sizes, and metrics. 
The SPEC harness is described in Section~\ref{sec:Harness}. 
The performance results follow in Sections~\ref{sec:Results}. The work related to performance benchmarking is reviewed in Section~\ref{sec:Rel-work}. The work is concluded in Section~\ref{sec:Conclusions}.

%% file: 02-motivation.tex
%
%
%
%

HPC application benchmarks are of great value for researchers and system managers who have used previous SPEC HPG benchmark suites, such as SPEC~MPI, SPEC~OMP, or SPEC~ACCEL, in various scenarios. 
The new SPEChpc 2021 benchmark suites also encourage similar uses.

\textbf{Procurement:}
HPC system procurement is an important flagship.
During the preparation of the tendering documents, managers compare SPEC scores across different hardware and software setups, available as submitted results online~\cite{specresults}. 
This helps define limits and thresholds in the tendering documents.
Given SPEC's application benchmark characteristics, the benchmark suites are highly valuable to extrapolate performance of complex scientific applications for various and especially future hardware architectures. 
To this end, it has been beneficial to further integrate SPEC scores into acceptance tests for HPC system procurement: 
vendors need to demonstrate that their products deliver high performance with scientific applications and not just on highly-optimized microbenchmarks. 
In the past, numerous organizations have extensively used SPEC HPG benchmarks for procurement purposes. 

\textbf{Research Software:}
Academic HPC systems provide users a varied HPC-based software stack, including research compilers and runtime systems. 
Academic researchers use the SPEC benchmarks to analyze whether the software stack is mature enough to compile and correctly execute these applications (using the verification feature of the SPEC harness), and whether the research compilers deliver high performance\mbox{~\cite{boehm2018evaluating,juckeland2016describing,pophale2019comparing,huber2018impact,cramer2021sxaurora,cperformance}}.
The SPEC HPG benchmarks have also been used during the prototype implementation of OMPT in LLVM's OpenMP runtime as part of the OpenMP tools committee work before the final specification of OMPT~\cite{OpenMPSpec50}.

\textbf{Performance Analysis of Tools:}
Development of software tools, such as performance analysis tools for parallel programs, is another HPC-related activity with high relevance for academic HPC systems.
The SPEC HPG benchmark suites are often used to measure the overhead of such tools, i.e., executing the SPEC benchmarks with and without the HPC software tool under development.
For example, the MUST tool provides runtime correctness and deadlock analysis of parallel programs.
MUST has been evaluated with SPEC~MPI~L2007~v2 to assess its overhead and the influence of specific changes in the tool infrastructure~\cite{hilbrich2012must}.
Similarly, specific parts of tools can be assessed using SPEC benchmarks, such as the OpenMP measurement adapters of Score-P~\cite{knuepfer2012scorep}, an instrumentation and measurement infrastructure for profiling and event tracing. 
The SPEC OMP benchmark suite has also been used to evaluate the existing Opari2 adapter against a prototype measurement adapter based on OMPT~\cite{Feld2019}.

\textbf{System Regression Testing:} 
HPC centers perform regular systems maintenance.
This includes security and software updates as well as performance optimizations. 
In the past, RWTH Aachen and TU Dresden observed that performance-relevant changes and errors arise unintentionally during this process.
%
Regression tests following maintenance intervals with well-defined SPEC benchmarks make it possible to detect such unintentional changes in the system.
The exact same application scenario is executed regularly, and the results are automatically compared against results from previous measurements. The development and testing of the SPEChpc 2021 benchmark suites and their execution on different HPC clusters have provided extremely useful information, especially for identifying non-performing HPC nodes. 
Two use cases encountered at sites RWTH Aachen and TU Dresden are described next.

\textbf{Case Study 1:} At RWTH Aachen University, tests with the new SPEChpc 2021 benchmark suites (\textit{medium} suite) on 50 compute nodes of the system showed significant negative performance differences for some of the benchmarks compared to other HPC systems with a very similar setup, e.g., available as SPEC results~\cite{specresults}.
Via a deep dive into the performance data, the execution times were found to differ mostly in the amount of MPI time, specifically, in MPI\_Allreduce collective operations.
This cross-node execution time imbalance is caused by: 
(1)~dropping memory bandwidth and (2)~system noise.
While the memory DIMMs did not completely malfunction and were not detected in the system's health check, some of them delivered roughly 20\,\% less bandwidth than expected. 
Hence, additional job-based bandwidth checks have been implemented and are regularly reported to the vendor responsible for replacing those DIMMs.
Although system noise is a known issue, its high impact for the worst-case scenarios as triggered by the SPEC benchmarks was surprising, underscoring the usefulness of the new benchmark suites.
To improve and achieve comparable performance for such bulk-synchronous parallel programs using MPI collectives, users at RWTH Aachen University will now be advised to leave one core empty per NUMA domain per compute node instead of fully occupying the nodes with MPI processes.
Given these important findings, the performance of other applications will also be investigated and, if necessary, improved.

\textbf{Case Study 2:} At TU Dresden, several undetected system problems were discovered on the test system B, even though system health checks were implemented.
The SPEChpc~2021 benchmarks revealed significant performance variations when executed multiple times across a changing subset of equal nodes of the system.  
The suites showed its high suitability to detect system problems (by means of comparison with reference data) that did not lead to a partial or complete failure of the system, but slowed down a subset of nodes significantly. 
Once the partial slow downs were recognized as such, their causes were found quickly to be: 
(1)~a faulty BIOS configuration of a number of computing nodes, 
(2)~a kernel bug~\cite{redhat2020kernelbug} occurring infrequently, and 
(3)~an unfavorable configuration of the SLURM daemon.
The commonality between all these issues is that no crashes occurred and the entire system was 100\% available at all times.
Nevertheless, benchmarks from the SPEChpc~2021 suites showed performance degradation of up to 50\% in the BIOS setup and kernel bug cases.

The above-mentioned kernel bug, was discovered due to the fact that the execution time of the Tealeaf benchmark (see Section~\ref{sec:Overview}) occasionally doubled inexplicably, even though the CPU resource configuration was unaltered.
Increasing the number of compute nodes also increased the frequency of observation of this phenomenon.

A profiler-assisted~\cite{knuepfer2012scorep} analysis of the core cycle counters over time revealed that a very small random fraction of MPI ranks were only processed at half speed because they received only half of the theoretically possible processor cycles.
This in turn led to execution time load imbalance of the actually very well (statically) balanced solver.
Deploying an additional system profiler~\cite{ilsche2017lo2s} revealed that errant kernel threads on the affected nodes were responsible for the absence of CPU cycles.
The root cause was determined to be a missing kernel patch~\cite{kernelpatch}, now installed.

%% file: 03-overview-benchmark.tex

The search program~\cite{specsearchprogram} for SPEChpc~2021 benchmarks from 2017 gathered applications from the HPC community with the following characteristics:
\begin{itemize}
    \item \textbf{Support for MPI+X parallelism}, where X takes one of three values: OpenMP host (denoted as OMP), OpenMP target (denoted as TGT), and/or OpenACC (denoted as ACC). They offer hybrid execution using all potentially available parallelism within and across compute nodes. An \textit{MPI-only} version of the application is needed for baseline comparisons. 
    \item \textbf{Origin in various science domains} to reflect the diversity of typical and real HPC workloads.
    \item \textbf{Fortran or C/C++} programming language.
    \item \textbf{Support for strong scaling} of work distribution for multiple data set sizes, as one SPEC suite will always distribute the same workload of a benchmark regardless of the actual number of process/threads used.
    \item \textbf{Predictable code paths} with no algorithmic differences depending on the computing platform
    \item \textbf{Limited time spent in I/O} as this not an I/O benchmark.
    \item \textbf{Numerically verifiable output} to check for correctness within a definable margin of error.
\end{itemize}






\subsection{SPEChpc~2021 Benchmark Composition}

\begin{table*}[!t]
    \centering
    \caption{SPEChpc~2021 benchmark application properties}
    \label{tab:BenchmarkApplications}
     \resizebox{18cm}{!}{%
    \begin{tabular}{|l|l|l|l|r|r|r|r|}
         \hline
         \textbf{Name} & \textbf{Application Area} & \textbf{Language} & \textbf{Suite} & \textbf{approx. \# LOC} & \textbf{\# MPI calls} & \textbf{\# OMP dir.} & \textbf{\# ACC dir.}\\
         \hline
         LBM D2Q37 (x05) & Computational Fluid Dynamics & C & T/S/M/L & 9,000 & 118 & 50 & 66 \\
         SOMA (x13) & Physics / Polymeric Systems & C & T/S/-/- & 9,500 & 90 & 192 & 185 \\
         Tealeaf (x18) & Physics, High Energy Physics & C & T/S/M/L & 5,400 & 22 & 86 & 40 \\
         Cloverleaf (x19) & Physics, High Energy Physics & Fortran & T/S/M/L & 12,500 & 23 & 827 & 886 \\
         Minisweep (x21) & Nuclear Engineering, Radiation Transport & C & T/S/-/- & 17,500 & 41 & 39 & 43 \\
         POT3D (x28) & Solar Physics & Fortran & T/S/M/L & (incl. HDF5) 495,000 & 88 & 124 & 77 \\
         SPH-EXA (x32) & Astrophysics and Cosmology & C++14 & T/S/-/- & 3,400 & 82 & 36 & 18 \\
         HPGMG-FV (x34) & Cosmology, Astrophysics, Combustion & C & T/S/M/L & 16,700 & 53 & 206 & 127 \\
         miniWeather (x35) & Weather & Fortran & T/S/M/L & 1,100 & 11 & 36 & 20 \\
        \hline
    \end{tabular}
    }
\end{table*}

The first official release (version 1.0.3) of the SPEChpc~2021 suites consists of \textit{nine} applications \cite{specresults}. 
The applications and their basic properties are summarized in Table~\ref{tab:BenchmarkApplications}.  
Not all application benchmarks are included in all the suites. 
This is the case when the maximum problem size that represents a realistic problem has already been reached in a smaller suite or when the problem does not naturally scale to all suites (see Section~\ref{subsec:suites}).
Other applications did not support all three levels of intra-node parallelism, namely `X' in MPI+X, before submission in response to the search program. 
During the benchmark preparation phase, those have been refactored accordingly to support the three `Xs'.
 


\subsection{SPEChpc~2021 Suites and Metrics}~\label{subsec:suites}
The SPEChpc~2021 benchmark suites support \textit{strong scaling} workloads by offering \textbf{four suites}: \textit{\textbf{tiny}}, \textit{\textbf{small}}, \textit{\textbf{medium}}, and \textit{\textbf{large}}, which represent common workload sizes for all benchmarks in one suite. 
Thus, the SPEC performance measurement harness (Section~\ref{sec:Harness}) can be used to execute and verify the results where everyone solves the same problem(s).

The \textbf{SPEC score} is the ratio of the execution time of the benchmarks on the reference system (RS) to the execution time on the system under test (SUT) that is also reported in the public SPEC results repository~\cite{specresults}. 
The \textit{reference performance timings} are generated on a cluster with Intel Haswell CPUs and Infiniband interconnection network using the MPI-only version of the benchmarks. 
The \textit{expected execution time} of one benchmark of a given suite on the reference system is around \textit{30 minutes}, usually achieved by adjusting  the benchmark's problem size and number of time iterations.

The maximum memory requirements for each workload size is defined to reflect typical HPC system sizes.
The benchmarks have also been tested with MPI rank counts typical of these system sizes. 
Due to communication buffers, the maximum memory requirements can easily be exceeded for very large MPI rank counts.
All these design limits are shown in Table~\ref{tab:DesignLimits}. 
Each suite is assigned a prefix number (Table~\ref{tab:DesignLimits}) and each benchmark a postfix number (Table~\ref{tab:BenchmarkApplications}). 
For example, \textbf{5}05 denotes the LBM benchmark in the \textit{\textbf{tiny}} suite. 
\begin{table}[htb]
    \centering
    \caption{Design limits for the SPEChpc~2021 suites}
    \label{tab:DesignLimits}
    \begin{tabular}{|l|r|r|r|}
        \hline
        \textbf{Suites} & \textbf{max memory} & \textbf{min ranks} & \textbf{max ranks}\\
        \hline
        \textit{tiny} (T) (5xx) & 64 GB & 1 & 256\\
        \textit{small} (S) (6xx)  & 480 GB & 64 & 1,024\\
        \textit{medium} (M) (7xx)  & 4 TB & 256 & 4,096\\
        \textit{large} (L) (8xx)  & 14.5 TB & 2,048 & 32,768\\
        \hline 
    \end{tabular}
\end{table}

%% file: 04-harness.tex
The SPEChpc~2021 suites reside within the SPEC harness. This harness is developed and has been maintained by SPEC for more than 15 years. This effort is shared across SPEC groups. 
Users of the popular SPEC CPU benchmarks\mbox{~\cite{bucek2018spec,spradling2007spec}} will be able to readily use SPEChpc~2021 benchmarks as well, as the harness used with both is the same and its usage very similar. 
The SPEC harness is involved in all aspects of running the benchmark, from installation, correctness to submission and publication of results~\cite{specresults}. 
It can be tuned to all usage scenarios, supports different compilers and compiler environments, as well as batch systems, code verification and ensures code source integrity. The harness supports publishing of benchmark results that include \textbf{all} the information required to reproduce the benchmark results. 
Therefore, the harness promotes performance reproducibility, which is of increasing importance to the HPC community.


%% file: 05-results.tex
We present here the results of executing the SPEChpc~2021 suites on four HPC systems: (a)~ Frontera at Texas Advanced Computing Center (TACC)~\cite{tacc},  (b)~Summit~\cite{summit} at ORNL,
(c)~JUWELS Booster module at Forschungszentrum J\"ulich~\cite{juwelsbooster}, and
(d)~Spock~\cite{spock}, a pre-exscale system also at ORNL. 
We have chosen these systems to show results and findings on a range of homogeneous (Frontera Intel Cascade Lake Xeon) and heterogeneous (Summit NVIDIA V100, JUWELS NVIDIA A100 and Spock AMD MI100) systems utilizing \textbf{MPI-only} and \textbf{MPI+X} programming paradigms.
All experiments used 1 MPI rank/GPU. 

We present \textbf{strong scaling} results as they typically indicate the best match between workload sizes and resources availability. 
The SPEC benchmark suites are traditionally evaluated using a SPEC score (defined in Section~\ref{sec:Harness}) which facilitates comparison between systems.
Given that performance comparisons of systems falls outside the scope of this work, we will report the traditional execution time in seconds. Table~\ref{tab:overviewExperiments} describes the experimental setup. 
We have also populated Zenodo~\cite{rawdata} with performance data that was generated to build the plots and tables in this manuscript. The results in Zenodo also entail performance comparisons of benchmarks across systems. The plots show that the comparisons may be interpreted in different ways drawing ambiguous conclusions, hence we do not focus on them in this paper but leave the interpretation of the comparisons to the different stakeholders including the procurement managers, HPC system operators, tools developers and application developers.

\begin{table}[t]
  \begin{center}
    \caption{Overview of the experimental setup}
    \label{tab:overviewExperiments}
    \resizebox{\columnwidth}{!}{%
    \begin{tabular}{|p{0.12\columnwidth}|p{0.25\columnwidth}|p{0.15\columnwidth}|p{0.07\columnwidth}|p{0.22\columnwidth}|}
    
      \hline
      \textbf{System} & \textbf{Dominant chips} & \textbf{Parallelism} & \textbf{Suite} & \textbf{min$\,\mid\,$max ranks}\\
      \hline
      Frontera & Intel Xeon Platinum 8280 & MPI-only, MPI+OMP & T, S, M, L & 56 $\mid$ 57,344 \\
      JUWELS Booster & NVIDIA A100 & MPI+ACC, MPI+TGT & M, L & 100 $\mid$ 1,400 \\
      Spock & AMD MI100 & MPI+TGT & S & 16 $\mid$ 32 \\
      Summit & NVIDIA V100 & MPI+ACC, MPI+TGT & M, L & 1,050 $\mid$ 16,800\\
      \hline
    \end{tabular}
    }
  \end{center}
 \vspace{-0.3cm}
\end{table}

\subsection{Performance Results on Frontera} 
\input{05-results-frontera}

\subsection{Performance Results on Summit}
\input{05-results-summit}

\subsection{Performance Results on JUWELS Booster}
\input{05-results-juwels}

\subsection{Performance Results on Spock}
\input{05-results-spock}

%% file: 05-results-frontera.tex
\begin{figure*}[p]
\centering
\begin{tabular}{cccc}
\rotatebox[origin=lB]{90}{\hspace{2cm}\small (a) \textit{Tiny} suite}
& 
\includegraphics[width=2.5in]{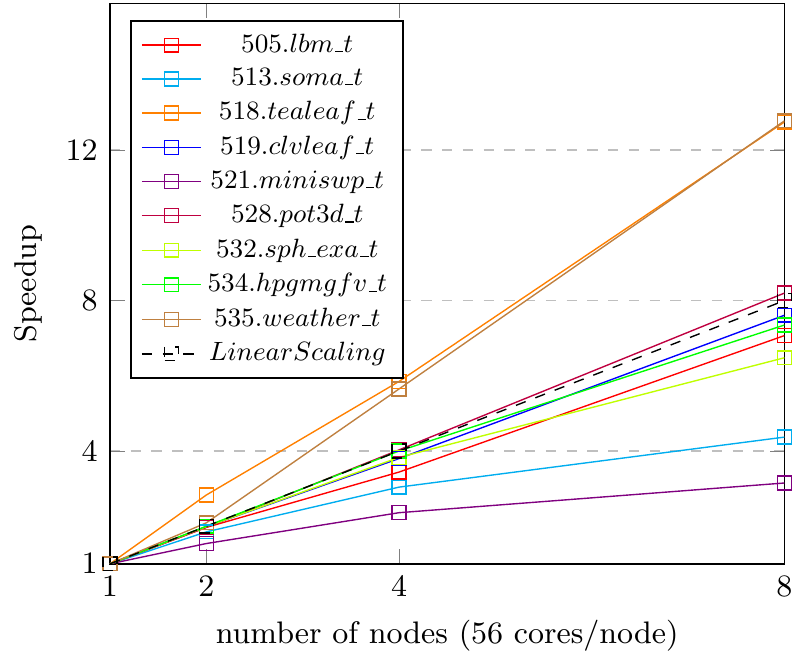}
&
\includegraphics[width=2.5in]{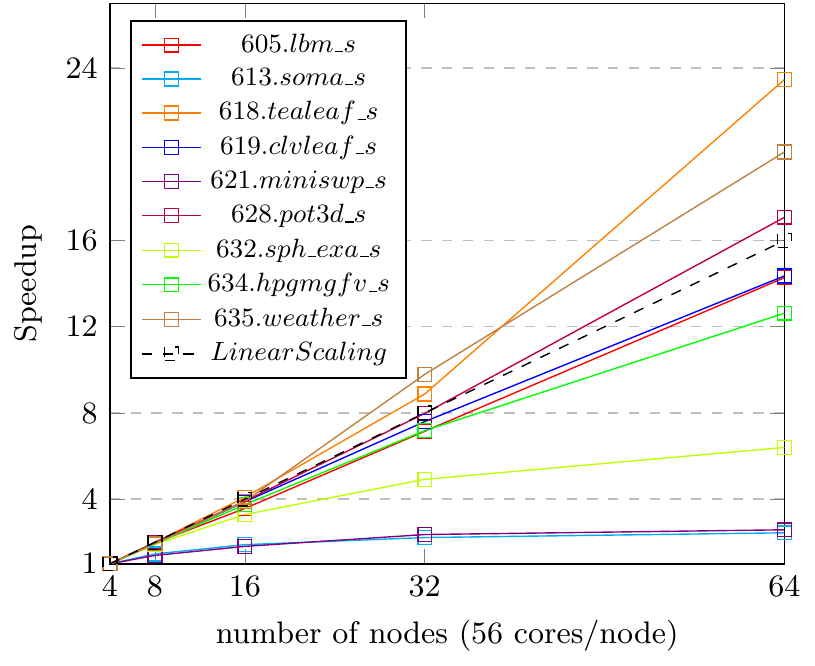}
    &
\rotatebox[origin=lB]{90}{\hspace{2cm}\small (b) \textit{Small} suite}
\\
\rotatebox[origin=lB]{90}{\hspace{2cm}\small (c) \textit{Medium} suite}
&
\includegraphics[width=2.5in]{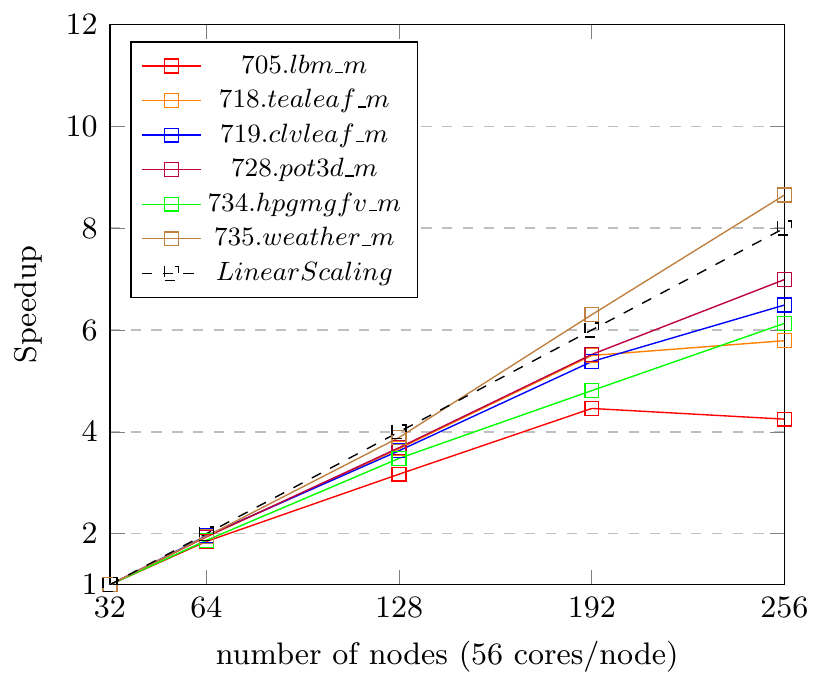}    
&
\includegraphics[width=2.5in]{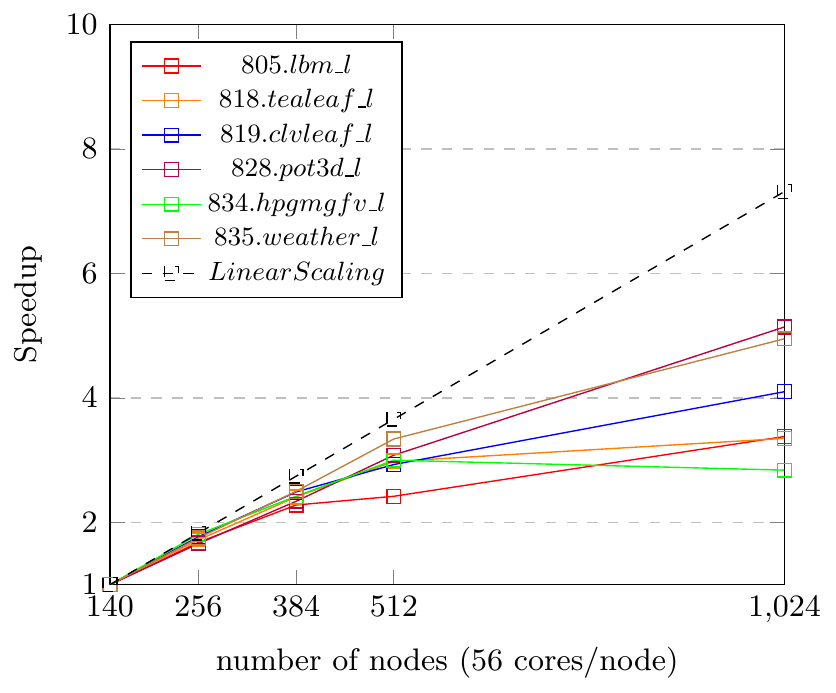}
&
\rotatebox[origin=lB]{90}{\hspace{2cm}\small (d) \textit{Large} suite}
\\
\end{tabular}
\caption{Speedup for MPI-only. \textit{Tiny}, \textit{Small}, \textit{Medium}, and \textit{Large} suites on Frontera}
\label{fig:frontera-mpi-scaling}
\end {figure*}

\begin{figure*}[p]
\centering
\begin{tabular}{cccc}
\rotatebox[origin=lB]{90}{\hspace{2cm}\small (a) \textit{Tiny} suite}
& 
\includegraphics[width=2.5in]{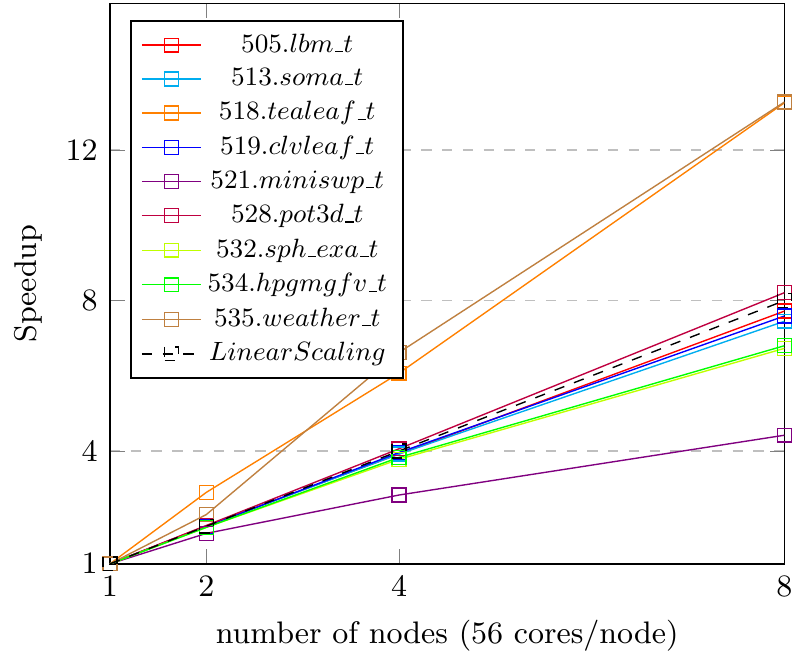}
&
\includegraphics[width=2.5in]{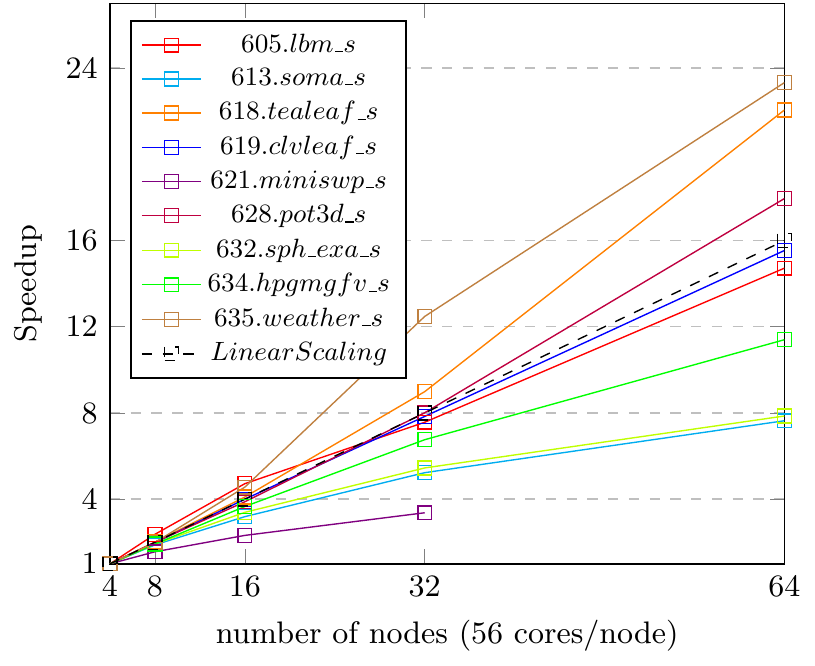}
    &
\rotatebox[origin=lB]{90}{\hspace{2cm}\small (b) \textit{Small} suite}
\\
\rotatebox[origin=lB]{90}{\hspace{2cm}\small (c) \textit{Medium} suite}
&
\includegraphics[width=2.5in]{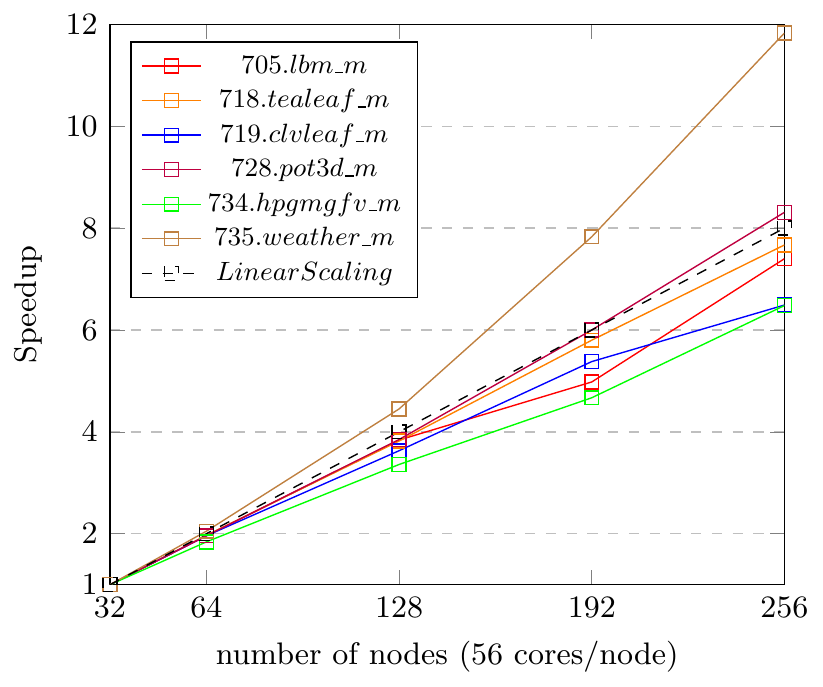}    
&
\includegraphics[width=2.5in]{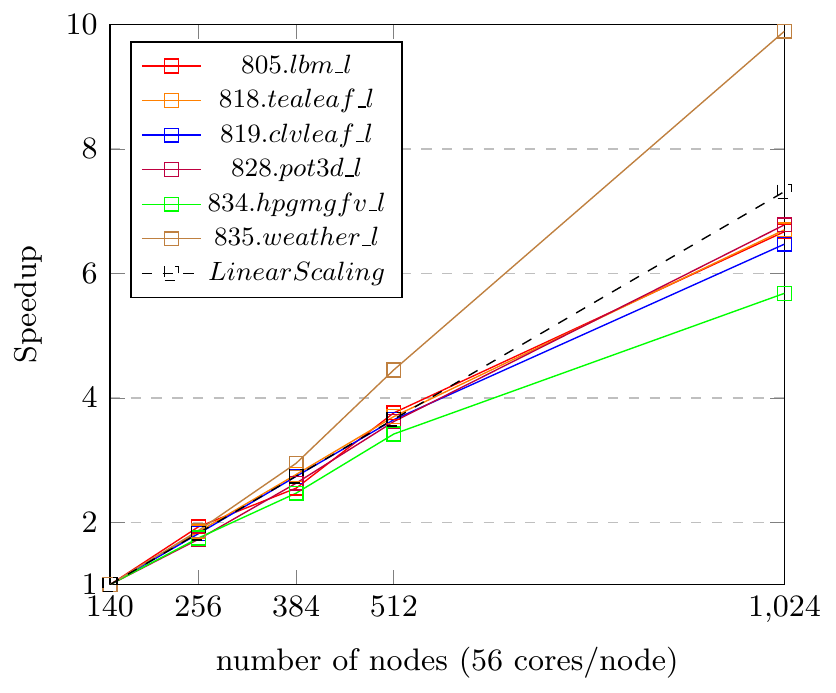}
&
\rotatebox[origin=lB]{90}{\hspace{2cm}\small (d) \textit{Large} suite}
\\
\end{tabular}
\caption{Speedup for MPI+OMP. \textit{Tiny}, \textit{Small}, \textit{Medium}, and \textit{Large} suites on Frontera. \textit{621.miniswp\_s} unable to run on 64 nodes due to a potential OpenMP compiler bug.}
\label{fig: frontera-omp-scaling}
\end {figure*} 


\begin{table}[ht]
  \begin{center}
    \caption{ Execution Time of the \textit{Large} Suite on Frontera$^\star$ \\ \centering(in seconds)  }
    \label{tab:frontera-timing-large}
    \resizebox{8.5cm}{!}{
    \begin{tabular}{|l|c|c|c|c|c|}
      \hline
       \textbf{Benchmark}     &\textbf{140 nodes} & \textbf{256 nodes} & \textbf{384 nodes} & \textbf{512 nodes} & \textbf{1024 nodes} \\
      \hline
      805.lbm\_l    & 998.6  & 517.1	& 391.2	& 265.9	& 149.6 \\
      818.tealeaf\_l & 828.1   & 448.2	& 298.6	& 223.7	& 123.6 \\
      819.clvleaf\_l & 1113.2  & 612.9	& 405.8	& 304.8	& 172.2 \\
      828.pot3d\_l   & 2593.0  & 1497.2	& 984.7	& 716.1	& 382.4 \\
      834.hpgmgfv\_l & 1045.4  & 596.9	& 423.8	& 305.6	& 183.9 \\
      835.weather\_l & 1207.5  & 644.2	& 408.9	& 271.3	& 122.1 \\
      \hline
    \end{tabular}
    }
  \end{center} 
  \begin{tablenotes}
\item $^\star$Results collected on Frontera for MPI+OpenMP executions
\end{tablenotes}
\end{table}

\begin{table}[htb]
    \centering
    \caption{SPEChpc~2021 Instruction Mix$^\star$  }
    \label{tab:vectorization}
     \resizebox{8.5cm}{!}{%
    \footnotesize
    \begin{tabular}{|l|c|c|c|c|}
         \hline
         \multirow{2}{*}{\textbf{Benchmark}} & \textbf{FP32 } & \textbf{FP64} & \textbf{Non-FP} & \textbf{Vectorization of FP} \\  
                   & \textbf{(\% of uOps)} & \textbf{(\% of uOps)} & \textbf{(\% of uOps)} & \textbf{ (\% of uOps)} \\  
 \hline
 605.lbm\_s & 0.00 & 51.98 & 48.02 & 86.80 \\
 613.soma\_s & 0.20 & 23.43 & 76.17 & 1.18 \\
 618.tealeaf\_s & 0.00 & 42.20 & 57.80 & 2.67 \\
 619.clvleaf\_s & 0.00 & 21.93 & 78.08 & 86.65 \\
 621.miniswp\_s & 0.00 & 8.92 & 91.07 & 57.90 \\
 628.pot3d\_s & 0.00 & 17.70 & 82.30 & 97.90 \\
 632.sph\_exa\_s & 0.00 & 36.27 & 63.70 & 49.75 \\
 634.hpgmgfv\_s & 0.00 & 22.30 & 77.70 & 81.22 \\
 635.weather\_s & 0.00 & 26.32 & 73.67 & 3.45 \\  
 \hline
    \end{tabular}
    } 
\begin{tablenotes}
\item $^\star$Results collected on Frontera for MPI-only executions using 4 nodes and 56 MPI ranks/node. Only results for the \textit{small} suite are shown; the codes in other suites exhibit nearly identical characteristics.
\end{tablenotes}
\end{table}

\begin{figure}[ht]
    \centering
    \subfigure[\textit{Small} suite]{
    \includegraphics[width=2.8in]{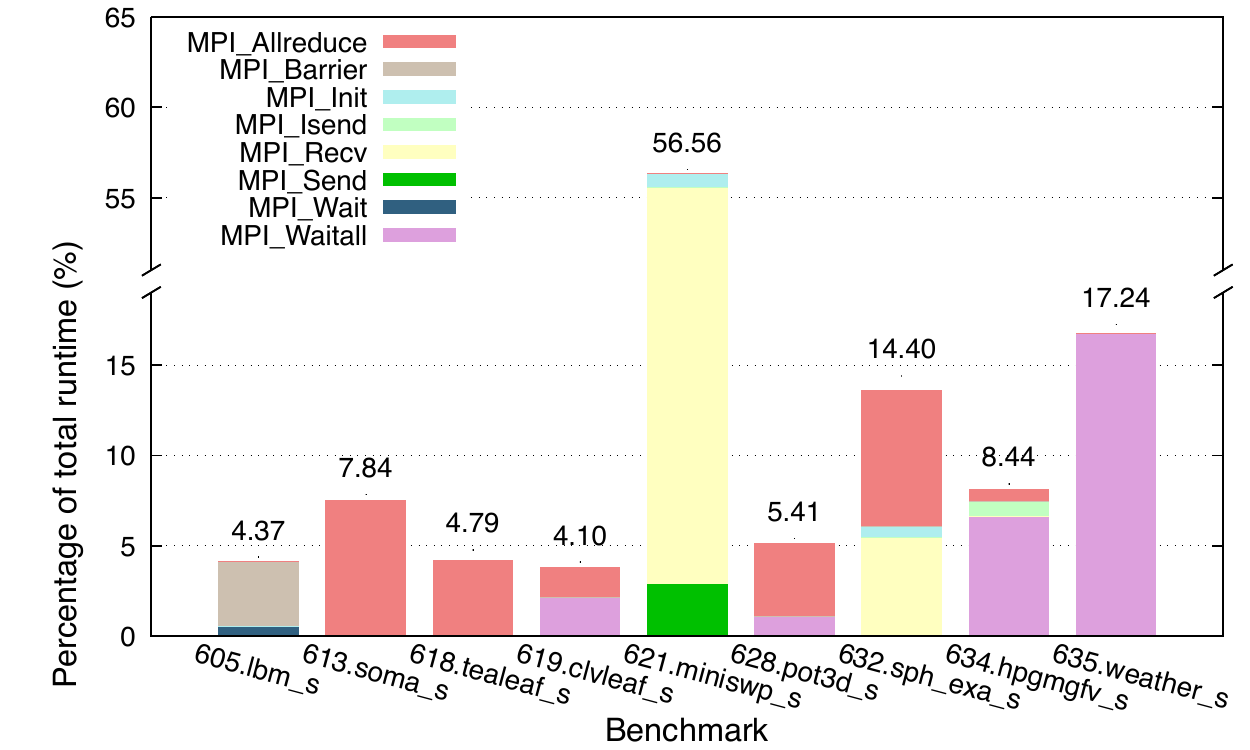}
    \label{fig:frontera_mpi_percent-small}
    } 
    \subfigure[\textit{Large} suite]{
    \includegraphics[width=2.8in]{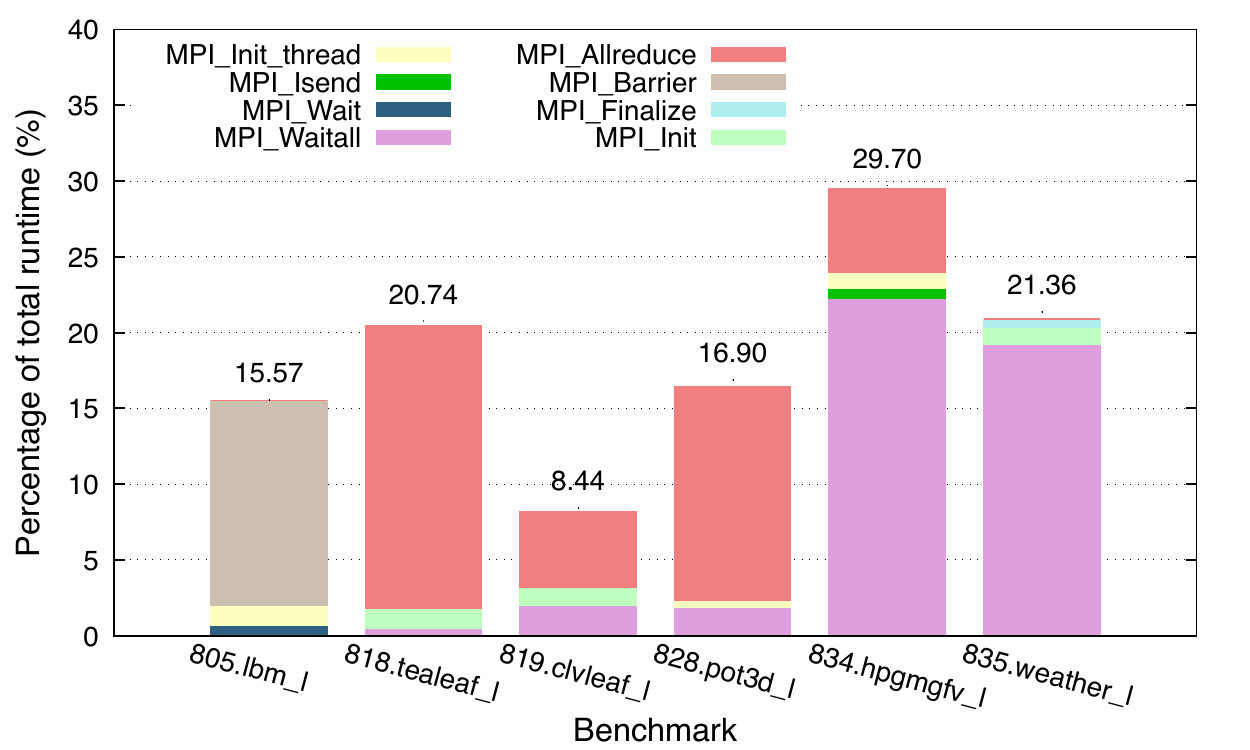}
    \label{fig:frontera_mpi_percent-large}
    } 
    \label{fig:frontera_mpi_percent} 
    \caption{MPI functions distribution for small and large suites. 
    Data collected on Frontera for MPI-only versions using 4 nodes and 56 MPI ranks/node for the \textit{small} suite, and 140 nodes and 56 MPI ranks/node for the large suite. Shown are only MPI functions that contribute more than 0.5\% of total execution time.}
\end{figure}

\begin{figure}[!t]
    \centering
    \includegraphics[width=2.8in]{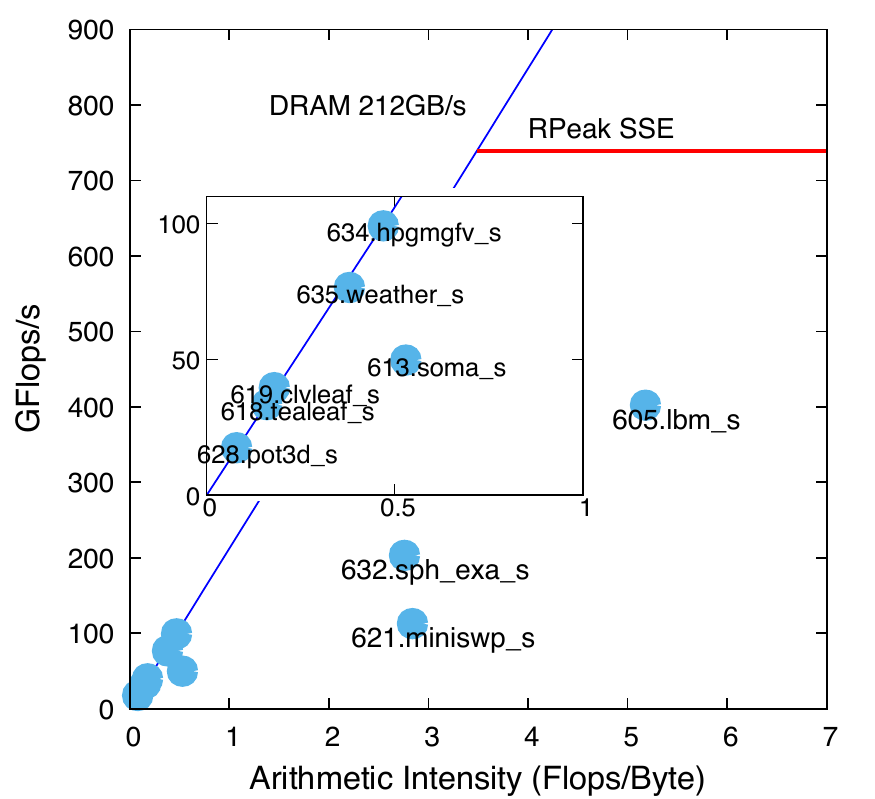}
    \caption{Roofline plot for the \textit{small} suite. Data collected for MPI-only versions using 4 nodes (224 ranks on Frontera). The roofline plots for the tiny, medium, and large suites are similar. Arithmetic intensity and memory bandwidth are collected for the entire duration of each program.}
    \label{fig:frontera-roofline}
\vspace{-0.4cm}
\end{figure}

\textbf{Experimental setup.}
Frontera is used to collect scalability and profiling data. 
The system consists of 8,368 Cascade Lake based Dell PowerEdge C6420 compute nodes with dual socket Intel Xeon Platinum 8280 28-core CPU and 192 GB DDR4 memory. 
The system's interconnection network is a fat tree topology with a blocking factor of 22:18. 
The interconnect is Infiniband HDR technology with full HDR connectivity between switches and HDR100 connectivity to the compute nodes.  
Since this is a CPU-only platform, MPI and MPI+OMP are the programming models used. 
For MPI-only executions, 56 MPI ranks per node were used. 
For MPI+OMP executions, 2 MPI ranks  with 28 OpenMP threads per node were used, each of the MPI ranks being placed on separate NUMA domains to minimize cross-NUMA OpenMP memory traffic. For timing data collection,  median is taken from 3 iterations of \textit{Tiny} and \textit{Small}, while \textit{Medium} and \textit{Large} were run for only 1 iteration. For all cases, the Intel Fortran/C/C++ compilers and Intel MPI from Intel Parallel Studio 2020 Update 4 were used. 
Codes are compiled with the \textit{-O3 -no-prec-div -fp-model fast=2 -xCORE-AVX512 -ipo} flags.  
All profiles were collected using Intel Application Performance Snapshot representing basic tuning. Fine tuning of MPI parameters, such as rank/thread distribution, can certainly further improve the performance.
Results on scalability and  performance statistics from the experiments on Frontera are next discussed. 

\textbf{Results.}\subsubsection{Scalability} To understand the parallel performance of the benchmark, a set of \textbf{strong scaling} tests were performed for all four suites with both MPI-only and MPI+OMP programming models. 
Fig.~\ref{fig:frontera-mpi-scaling} shows MPI-only scalability, and Fig.~\ref{fig: frontera-omp-scaling} demonstrates the MPI+OMP scalability. The execution times of the large suite with MPI-only are presented in Table~\ref{tab:frontera-timing-large}. For additional information on execution times of other suites, please refer to performance data collected via Zenodo available at~\cite{rawdata}. 

For the MPI-only runs, all suites scale well within their design limit, and have their appropriate applicable ranges. 
From the \textit{tiny} suite, Minisweep and SOMA scale relatively poorly and scaling efficiency drops below 50\% when more than 4 nodes are used. 
For SOMA, this is due to a high volume of all-to-all communication which is the nature of the code.  
Regarding Minisweep, the poor scaling beyond certain node count is due to a combination of inherently high amount of MPI communication and relatively small data set being used to fit the designed memory limit of the benchmark. 
As soon as the \textit{tiny} suite no longer scales well, the \textit{small} suite should be considered unless benchmarking MPI and interconnect is the main goal.   
From the \textit{small} suite, Minisweep and SOMA also scale relatively poorly compared to other codes for the same reasons outlined above. 
Another code, SPH-EXA, joins the list of non-ideal scaling at 32 nodes also due to high volume of MPI traffic and reduced compute work per rank.  
Beyond 32 nodes on Frontera, the \textit{medium} suite works best on up to 192 nodes, beyond which the \textit{large} suite should be favored.  

In the hybrid MPI+OMP case, performance is significantly better for several codes due to reduced MPI all-to-all communication and/or reduced memory traffic given the shared memory model. 
It is also worth mentioning that two codes, Tealeaf and miniWeather, will often achieve super-linear scaling. 
These two codes have the smallest memory footprint in the suites, yet are highly memory-bound. 
As more nodes are used to solve the same problem, the problem size per node reduces, resulting in decreased memory traffic and fewer memory stalls. 

For example, miniWeather in the \textit{small} suite (635.weahter\_s) used 220.03 GB/s as peak memory bandwidth and 202.05 GB/s as average bandwidth when 4 nodes are used; this is close to the upper limit of the achievable memory bandwidth, while at 16 nodes, the memory bandwidth utilization was reduced to 197.52 GB/s as peak and 146.44 GB/s as average. 
On 32 nodes, the metrics further reduced to 155.29 GB/s and 118.11 GB/s, respectively. 
The resulting DRAM stall was 33.43\%, 11.27\%, 7.31\% at 4, 16, and 32 nodes, respectively.  
Such a change in performance characteristics alleviates performance bottleneck and yields the super-linear scaling behavior. 

Demonstrated by all the scalability results, with the four different suites, the SPEChpc~2021 benchmark covers the entire spectrum from a few nodes to few hundreds. The \textit{tiny} suite is suitable for a single node or a few nodes, the \textit{small} suite is bigger and will be best suited to test a handful of nodes with a few hundred cores. The \textit{medium} suite works well on a small cluster with a few thousand cores, while the \textit{large} suite is large enough to test a medium-sized cluster with tens of thousands cores.   

\subsubsection{Performance statistics}
For an MPI involved benchmark, analysis of high level statistics of MPI calls is of interest. 
MPI profiles were collected for all the codes in all four suites. 
In this manuscript, profiles of the \textit{small} suite executed on 4 nodes (224 ranks), representing the performance characteristics at small scale, and of the \textit{large} suite executed on 140 nodes (7840 ranks), illustrating the performance characteristics at large scale, are shown. For additional information please refer to performance data collected via Zenodo available at~\cite{rawdata}. 
The MPI functions that consume more than 0.5\% of total execution time are shown in Fig.~\ref{fig:frontera_mpi_percent-small} and Fig.~\ref{fig:frontera_mpi_percent-large} for the \textit{small} and \textit{large} suites, respectively.  
Clearly, \textit{MPI\_Allreduce} plays a big role in many codes, such as SOMA, Tealeaf, Cloverleaf, POT3D, and SPH-EXA. 
Point-to-point communications are a key component for Minisweep, SPH-EXA, HPGMG-FV, and miniWeather. 
The purple bar denotes \textit{MPI\_Waitall} and indicates codes that rely on large amount of non-blocking communication.  

In addition, the floating point metrics were inspected.
The instruction mix of single precision floating point (FP32) operations, double precision floating point (FP64) operations, and non-floating point (non-FP) operations, along with the SIMD vectorization rate are shown in Table \ref{tab:vectorization}. 
The statistics are identical for different workloads; thereby, only the data for the \textit{small} suite are presented here. 
As the amount of AVX512 or AVX2 vectorization depends on the compiler switches used, we only show the total percentage of FP operations being vectorized and do not distinguish the underlying instruction set.  
These codes have a healthy mix of FP and non-FP operations, and most codes have their FP operations very well vectorized.  
The codes are FP64-heavy, with only SOMA having a tiny percentage of FP32 operations. Memory bandwidth is the limiting factor for many HPC codes. Most SPEChpc~2021 codes are also memory-bound.  
The roofline analysis in Fig. \ref{fig:frontera-roofline} reveals that five out of the nine codes land on the DRAM bandwidth boundary. 
This is for the \textit{small} suite executed on 4 nodes (224 ranks); other workloads executed at the lower range of the design limits (Table~\ref{tab:DesignLimits}) show very similar characteristics. 
As more resources are added, codes like TeaLeaf and miniWeather become less memory-bound as mentioned in the above super-linear scaling discussion.  
All floating point operations of the the most compute-intensive code, LBM, in the suite can be vectorized; this code will benefit most from a long SIMD instruction set. 

%% file: 05-results-summit.tex


\begin{figure}
    \centering
    \includegraphics[width=2.5in]{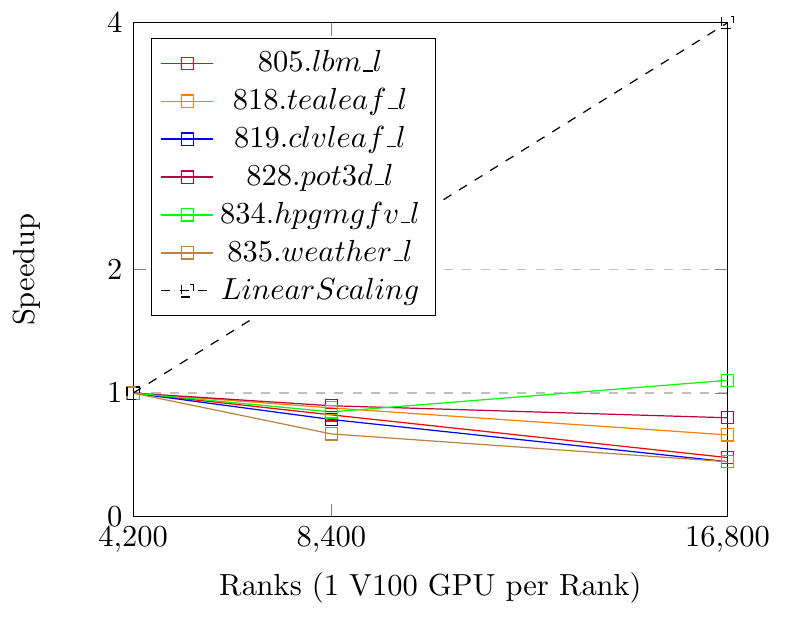}
    \caption{Speedup for MPI+ACC, \textit{Large} Suite on Summit}
    \label{fig:OpenACC-Summit-Large} 
\vspace{-2mm}
\end{figure}
\begin{figure}
    \centering
    \includegraphics[width=2.5in]{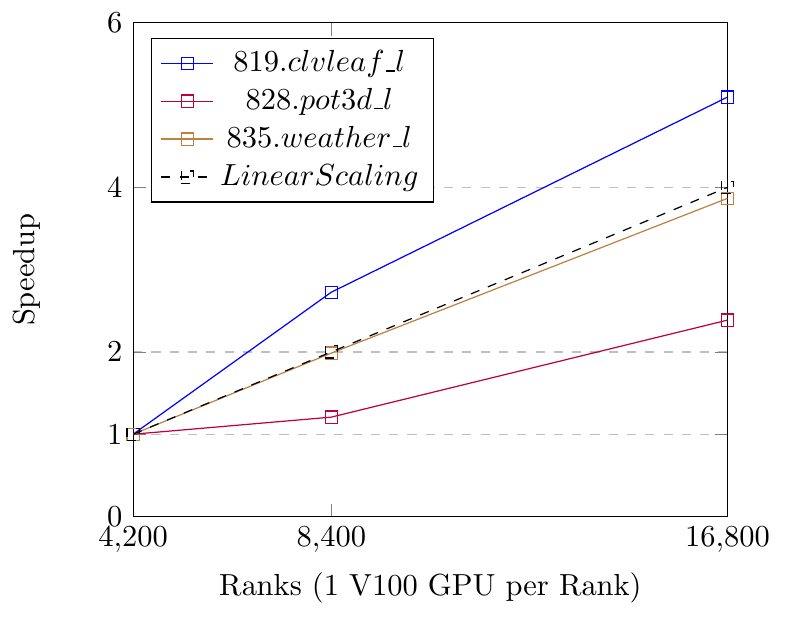}
    \caption{Speedup for MPI+TGT, \textit{Large} Suite on Summit}
    \label{fig:OpenMP-Summit-Large} 
\end{figure}
\begin{figure}
    \centering
    \includegraphics[width=2.5in]{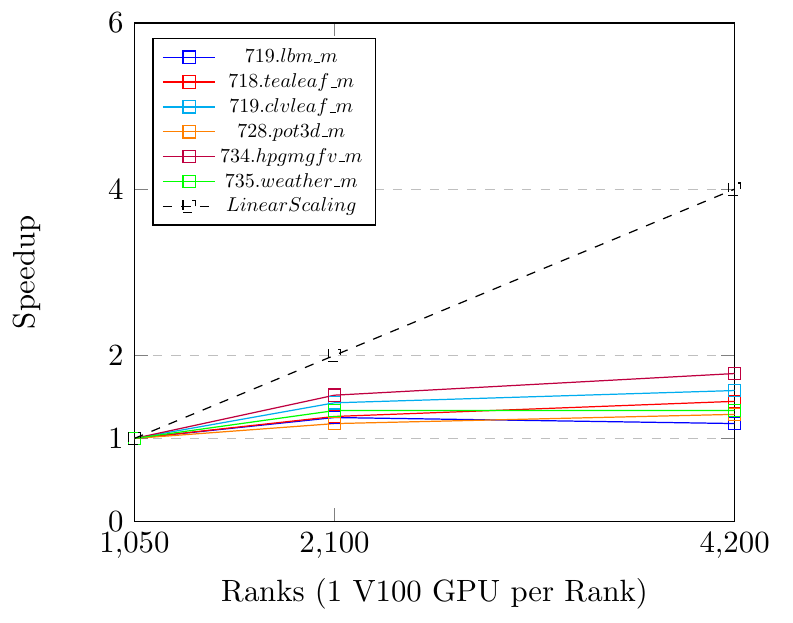}
    \caption{Speedup with MPI+ACC, \textit{Medium} Suite on Summit}
     \label{fig:OpenACC-Summit-Medium} 
 \vspace{-2mm}
\end{figure}
\begin{figure}
    \centering
    \includegraphics[width=2.5in]{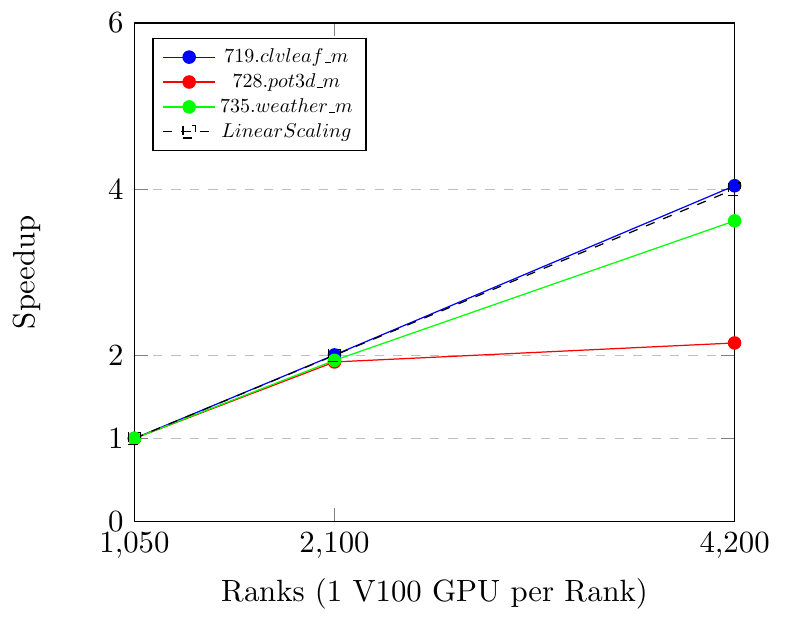}
    \caption{Speedup with MPI+TGT, \textit{Medium} Suite on Summit}
     \label{fig:OpenMP-Summit-Medium} 
\end{figure}

\begin{table}[ht]
  \begin{center}
    \caption{ Execution Time of the \textit{Medium} and  \textit{Large} \\ \centering Suites on Summit  (in seconds) }
    \label{tab:summit_medium_large}
    \resizebox{7cm}{!}{%
    \begin{tabular}{|l|r|r|r|r|r|r|}
      \hline
      \multicolumn{7}{|c|}{\textbf{Medium Suite}} \\
      \hline
      \multicolumn{1}{|c|}{\textbf{Benchmark}} & \multicolumn{2}{|c|}{\textbf{1050 Ranks}} & \multicolumn{2}{|c|}{\textbf{2100 Ranks}} & \multicolumn{2}{|c|}{\textbf{4200 Ranks}} \\
      \hline
       & \textbf{ACC} & \textbf{TGT} & \textbf{ACC} & \textbf{TGT} & \textbf{ACC} & \textbf{TGT} \\
      \hline
      705.lbm\_m     & 20.4 & CE & 16.3 & CE & 17.3 & CE  \\
      718.tealeaf\_m & 71.3 & RE & 56.4 & RE & 49.3 & RE  \\
      719.clvleaf\_m & 32.0 & 4154 & 22.4 & 2072 & 20.3 & 1028 \\
      728.pot3d\_m   & 95.7 & 3159 & 81.2 & 1645 & 74.2 & 1470 \\
      734.hpgmgfv\_m & 187 & RE & 123 & RE & 105 & RE \\
      735.weather\_m & 27.1 & 5416 & 20.3 & 2792 & 20.3 & 1497 \\
      \hline

      \multicolumn{7}{|c|}{\textbf{Large Suite}}\\
      \hline
      \multicolumn{1}{|c|}{\textbf{Benchmark}} & \multicolumn{2}{|c|}{\textbf{4200 Ranks}} & \multicolumn{2}{|c|}{\textbf{8400 Ranks}} & \multicolumn{2}{|r|}{\textbf{16800 Ranks}} \\
      \hline
       & \textbf{ACC} & \textbf{TGT} & \textbf{ACC} & \textbf{TGT} & \textbf{ACC} & \textbf{TGT}\\
      \hline
      805.lbm\_l     & 31.8 & CE & 38.6 & CE & 66.6 & CE \\
      818.tealeaf\_l & 60.1 & RE & 68.3 & RE & 90.9 & RE \\
      819.clvleaf\_l & 29.4 & 3415 & 37.4 & 1253 & 66.2 & 670 \\
      828.pot3d\_l   & 140 & 3618 & 156 & 2996 & 175 & 1516 \\
      834.hpgmgfv\_l & 128 & RE & 151 & RE & 116 & RE \\
      835.weather\_l & 26.3 & 4887 & 39.3 & 2461 & 59.2 & 1264 \\
      \hline
    \end{tabular}
    }
  \end{center}
\vspace{-4mm}
\end{table}

\textbf{Experimental setup.} These results use ORNL's Summit supercomputer which consists of over 4,600 nodes each with two 22-core IBM POWER9 CPU and six NVIDIA V100 GPUs. 
On Summit, the MPI+ACC and MPI+TGT models were executed using NVHPC 21.7 and IBM XL 16.1.1-10, respectively. For MPI+ACC, the benchmarks were compiled using \textit{-O3 -acc=gpu}. For MPI+TGT, the benchmarks were compiled using \textit{-O3 -qarch=pwr9 -qtune=pwr9 -qsmp=omp -qoffload -qtgtarch=auto}.
For the experiments using MPI+ACC at-scale, two iterations were used. For MPI+TGT, however, due to the long execution time, a single iteration was used. All experiments on Summit used 1 MPI rank per V100 GPU. 

\textbf{Results.}
The experiments on Summit use both MPI+ACC and MPI+TGT programming models. 
 Fig.~\ref{fig:OpenACC-Summit-Large} shows that running the \textit{large} suite with more than 4,200 MPI ranks, each offloading to a single V100 GPU, results in poor scalability. 
The problem sizes in the \textit{large} suite likely need to be increased to accommodate systems of Summit's scale. 
Only three of the six benchmarks successfully compiled with IBM XL 16.1.1-10 compiler. 
As shown in Fig.~\ref{fig:OpenMP-Summit-Large}, although performance improves as the number of MPI ranks increases, overall, the execution time is over 200x slower than that obtained with MPI+ACC using the same number of GPUs (see Table~\ref{tab:summit_medium_large}). An important point to take away is that performance largely depends on the implementations. 
In addition, performance of the \textit{medium} suite was investigated using 1,050 to 4,200 ranks, each offloading to a single V100 GPU. Both programming models at this scale show increasing speedup. However, as shown in Table~\ref{tab:summit_medium_large}, the execution time in these cases is 30-200X slower. Additional studies are necessary to understand this degradation. 
Fig.~\ref{fig:OpenACC-Summit-Medium} and Fig.~\ref{fig:OpenMP-Summit-Medium} show strong scaling results for the \textit{medium} suite.

%% file: 05-results-juwels.tex
\begin{table}[h!]
  \begin{center}
    \caption{Execution time of the \textit{Large} Suite on JUWELS \\ \centering  Booster  (in seconds)}
    \label{tab:juwels_table_large}
    \resizebox{7cm}{!}{%
    \begin{tabular}{|l|r|r|r|r|r|r|}
      \hline
      \multicolumn{1}{|c|}{\textbf{Benchmark}} & \multicolumn{2}{|c|}{\textbf{400 Ranks}} & \multicolumn{2}{|c|}{\textbf{800 Ranks}} & \multicolumn{2}{|r|}{\textbf{1400 Ranks}} \\
      \hline
       & \textbf{ACC} & \textbf{TGT} & \textbf{ACC} & \textbf{TGT} & \textbf{ACC} & \textbf{TGT}\\
      \hline
      805.lbm\_l & 56.7 & 70.9 & 32.5 & 40 & 22.2 & 26.6 \\
    818.tealeaf\_l & 154 & 70.7 & 106 & 53.5 & 73.4 & 38.8 \\
    819.clvleaf\_l & 61.9 & 236 & 33.6 & 128 & 20.4 & 84.2 \\
    828.pot3d\_l   & 200 & 280 & 150 & 205 & 95 & 136 \\
    834.hpgmgfv\_l & 209 & 539 & 141 & 332 & 133 & 229 \\
    835.weather\_l & 59.9 & 73.4 & 37.4 & 48.3 & 23.7 & 34.1   \\
      \hline
    \end{tabular}
    }
  \end{center}
   \vspace{-2mm}
\end{table}

\textbf{Experimental setup.}
The JUWELS Booster module consists of NVIDIA's A100 GPUs and AMD EPYC Rome CPUs. 
For both MPI+TGT and MPI+ACC runs, the GCC 10.3.0 and NVHPC 21.5 compilers are used, respectively, while ParaStation MPI/5.2.9-1 is used for MPI. 
Each of the JUWELS booster module consists of 4 NVIDIA A100 GPUs. For both MPI+ACC and MPI+TGT versions, the benchmarks are compiled using \textit{-w -Mfprelaxed -Mnouniform -Mstack\_arrays -fast}. 
For all the experiments, due to long execution time, one iteration was used. 

\begin{figure}
    \centering
    \includegraphics[width=2.5in]{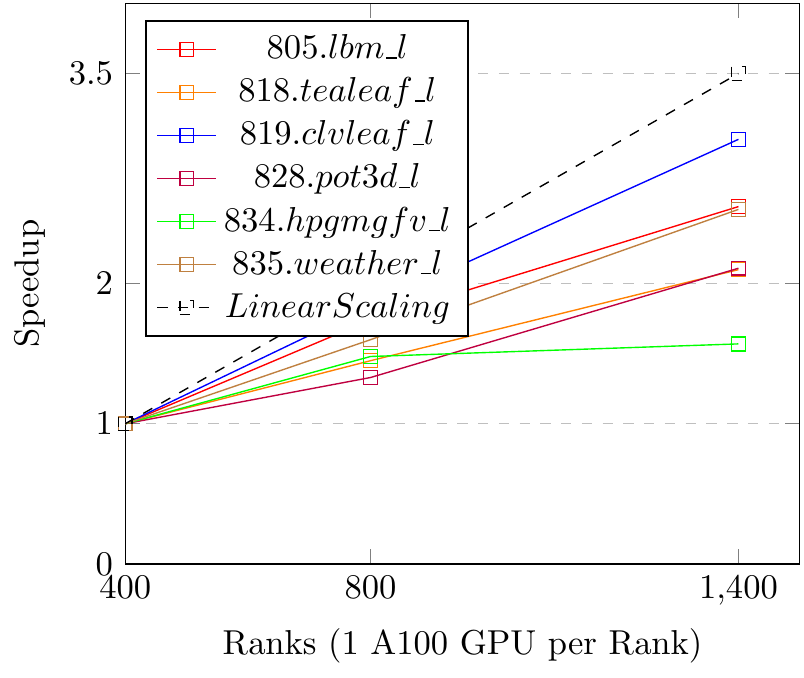}
    \caption{Speedup for MPI+ACC, \textit{Large} Suite on JUWELS Booster}
     \label{fig:juwels-acc-scaling-large} 
 \vspace{-4mm}
\end{figure}

\begin{figure}
    \centering
    \includegraphics[width=2.5in]{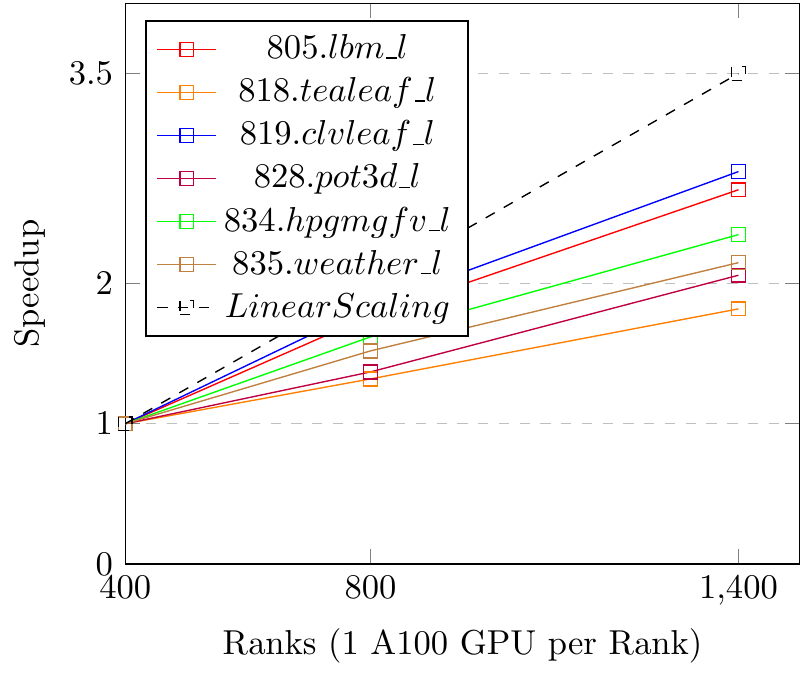}
    \caption{Speedup for MPI+TGT, \textit{Large} Suite on JUWELS Booster} 
     \label{fig:juwels-omp-scaling-large} 
 \vspace{-4mm}
\end{figure}

\textbf{Results.}
The \textbf{strong scaling} results on JUWELS for the \textit{large} suite using both MPI+ACC and MPI+TGT versions are shown in Figures~\ref{fig:juwels-acc-scaling-large} and \ref{fig:juwels-omp-scaling-large}. These 
show that almost all six benchmarks achieve close to linear scaling with room for improvement. 
For both MPI+ACC and MPI+TGT versions, Cloverleaf shows the best scaling behavior followed by LBM and the rest. 
Table~\ref{tab:juwels_table_large} presents the execution time in seconds for the \textit{large} suite for both versions spanning 400 to 1,400 MPI ranks. 

We observe that for LBM, Cloverleaf, POT3D, \mbox{HPGMG-FV} and miniWeather, the MPI+ACC version performs better than the MPI+TGT version. Cloverleaf is evidently over 3.9x faster on an average across ranks and HPGMG-FV over 2.2x faster on an average across ranks. However for Tealeaf, the MPI+TGT version is about 2x better than the MPI+ACC version. 
The observations above were similar for the \textit{medium} suite as well. Future work will include investigation of these discrepancies in more detail. For additional information on any of these experiments along with data on the \textit{medium} suite, please refer to performance data collected in Zenodo~\cite{rawdata}. 

While comparing the parallel efficiency of the \textit{large} suite with that of the \textit{medium} suite, we observe that the efficiency of medium
workload drops below 75\% with 400 ranks for almost all the benchmarks, indicating that for ranks above 400, it is better to use larger workload to achieve the best parallel efficiency especially on large clusters such as JUWELS. The trend seems to hold for both MPI+ACC and MPI+TGT versions.

%% file: 05-results-spock.tex
\begin{table}[!htb]
  \begin{center}
    \caption{Execution Time of the \textit{Small} Suite on \\ \centering  Spock  (in seconds) }    
    \label{tab:spock_small}
    \resizebox{7cm}{!}{%
    \begin{tabular}{|l|r|r|r|}
      \hline
      \textbf{Benchmark} & \textbf{16 Ranks} & \textbf{24 Ranks} & \textbf{32 Ranks} \\
      \hline
      605.lbm\_s & 234.2 & 169.4 & 142.3 \\
      618.tealeaf\_s & 624.2 & 507.6 & 461.8 \\
      621.miniswp\_s & 1319.6 & 967.9 & 828.1 \\
      628.pot3d\_s & 1426.1 & 968.7 & 897.4 \\
      632.sph\_exa\_s & 1065.5 & 747.2 & 631.6 \\
      635.weather\_s & 5154.7 & 3305.4 & 2226.4 \\
      \hline
    \end{tabular}
    }
  \end{center}
\vspace{-4mm}
\end{table}

\begin{figure}
    \centering
    \includegraphics[width=2.5in]{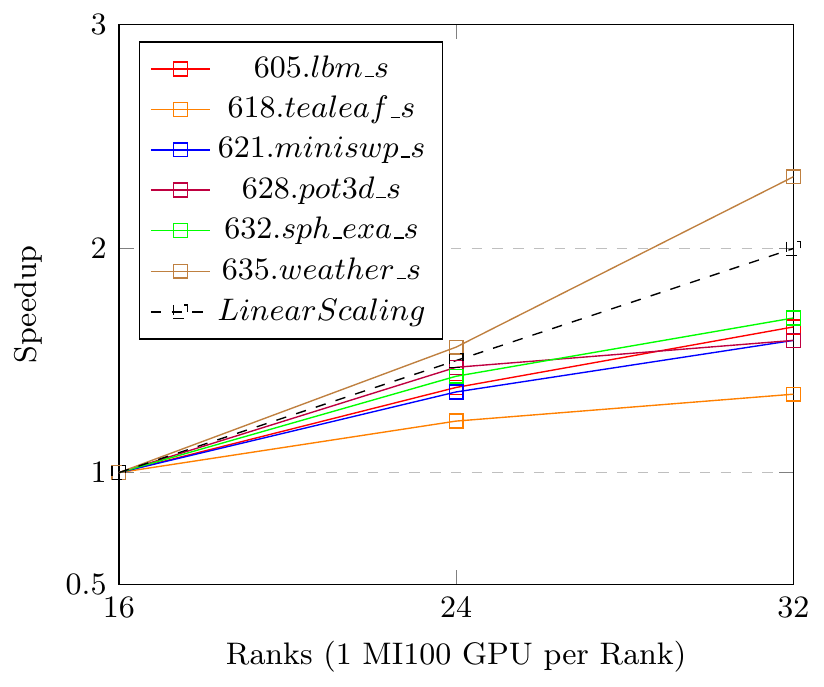}
    \caption{Speedup for MPI+TGT, \textit{Small} Suite on Spock}
    \label{fig:OpenMP-TGT-Spock-Small} 
 \vspace{-4mm}
\end{figure}

\textbf{Experimental setup.}
The Spock pre-production AMD~\cite{spock} system at ORNL is used to collect scalability and profiling data using the SPEChpc~2021 benchmark suites. Spock consists of 36 compute nodes, each with one AMD EPYC 7662 64-Core processor, 256 GB DDR4 memory, and four AMD Instinct MI100 GPUs.
These experiments used 16, 24, and 32 MPI ranks with 4 ranks per node corresponding to 4, 6 and 8 nodes. Two iterations were used for the 4- and 8-node runs and 3 iterations were used for the 6-node run.
The reported time is the average over all iterations.
The benchmark suites were compiled with the LLVM Clang v12.0.0 compiler suite packaged with ROCm v4.2.0, acceleration libraries for the MI100 architecture, and Cray MPICH v8.1.8.
The benchmarks were compiled using \textit{-O3 -fopenmp -target x86\_64-pc-linux-gnu -fopenmp-targets=amdgcn-amd-amdhsa -Xopenmp-target=amdgcn-amd-amdhsa -march=gfx908}.

\textbf{Results.} We ran SPEChpc~2021 \textit{small} suite on Spock.
SOMA could not compile because of an internal compiler error while processing the TGT directives.
HPGMG-FV did not run to completion within the time allowed for jobs on Spock; this is potentially caused by the TGT or MPI implementations.
The exact cause is unknown and investigation is ongoing.
We present \textbf{strong scalability} plots for the \textit{small} suite on Spock in Fig.~\ref{fig:OpenMP-TGT-Spock-Small}.
The execution times for these codes (shown in Table~\ref{tab:spock_small}) decreased for all codes as more nodes were added.

%% file: 07-related.tex
Numerous benchmarks focus on hardware performance features.
For instance, SHOC~\cite{danalis2010scalable} that includes both the OpenCL and CUDA implementations provides microbenchmarks and small application kernels.
The classic NASA parallel benchmarks (NPB)~\cite{bailey1995parallel} are a popular multi-zone, hybrid (MPI+X) benchmark suite~\cite{wu2011performance,jin2006performance} that are used for evaluating prototypes of programming features, runtime or compiler implementations. 
The Edinburgh Parallel Computing Centre (EPCC) developed a benchmark suite that consists of a set of low-level operations designed to test raw performances of compilers and hardware~\cite{bull2001microbenchmark}. 
The HPC community also widely uses HPL~\cite{dongarra2003linpack} for floating-point focused operations, HPMG~\cite{adams2014hpgmg} for multigrid methods, and HPCG~\cite{heroux2013hpcg} for bandwidth-focused operations.
The benchmarking landscape has also been influenced by so-called mini or proxy applications that abstract a specific (performance) behaviour from the real-world application they mimic. 
Examples are Mantevo~\cite{heroux2009improving} and Exascale Computing Project's (ECP) Mini-apps~\cite{cook2017proxy,dobrev2017ecp,sultana2018understanding}. 
The SPEChpc~2021 suites differ from the above benchmarking effort in a way that the suites are incorporated into the SPEC harness (detailed in Section~\ref{sec:Harness}) adding benefits such as reproducibility, results publication, and ranking of systems utilizing the SPEC scores. 
Other SPEC related benchmarking efforts include the release of the following suites for the HPC community: 
SPEC HPC96~\cite{eigenmann1996benchmarking} benchmark suites, improved and replaced by SPEC HPC2002~\cite{eigenmann2002spec}, 
SPEC OMP2001~\cite{eigenmann-omp2001}, superseded by SPEC~OMP2012~\cite{muller2012spec} for shared-memory parallel systems using CPUs,
SPEC~MPI2007~\cite{muller2010spec} for distributed-memory parallel systems using CPUs, and
SPEC~ACCEL~\cite{juckeland2014spec} with applications using OpenACC and OpenMP 4.5 in a performance portable manner, thus, targeting different (single) accelerator devices.
SPEChpc~2021 suites complement the above SPEC suites by enabling scalability and efficiency studies at large scale, using large homogeneous and heterogeneous HPC clusters that are recently increasingly being built with multiple-GPUs per node.

%% file: 08-conclusion-future-work.tex
This work illustrates first experiences in performance benchmarking with the new SPEChpc~2021 suites on 4 HPC systems: Frontera, Summit, JUWELS Booster, and Spock, that have diverse hardware architectures: Intel Cascade Lake CPUs, NVIDIA V100/A100 GPUs, and AMD Instinct MI100 GPUs. 
The SPEChpc~2021 suites (tiny, small, medium, and large) employ MPI-only, MPI+OMP, MPI+ACC, and MPI+TGT as parallel programming APIs.
Analysis of SPEChpc's application properties on Frontera reveals that 
(1)~most codes are memory-bound, 
(2)~they utilize a good instruction mix, and 
(3)~8-30\% of the total execution time of the~\textit{large} suite corresponds to MPI, in particular \textit{MPI\_Allreduce}. 
This analysis forms the basis for future research using the SPEChpc~2021 suites. 

We presented strong-scaling results for all suites on all systems with up to 16,800 MPI ranks. 
Selecting an appropriate suite for performance benchmarking is critical for trading off poor scalability with superlinearity. 
The best and the worst parallel efficiencies were obtained on the \mbox{state-of-the-art} NVIDIA’s A100 GPUs. 
On the same accelerator-based systems, the performance of MPI+TGT implementations lags behind that of MPI+ACC implementations. 
These performance differences are due to the OpenMP offloading features' implementations in compilers (e.g., NVHPC vs IBM XL vs LLVM) which require further in-depth investigation. 
These results have been motivating compiler vendors to invest efforts towards closing the performance gaps.
All performance measurements presented here will be published on the SPEC website~\cite{specresults} upon the official release of the SPEChpc~2021 suites by SPEC~HPG.
This activity is ongoing, while SPEC HPG finalizes the detailed review criteria and documentation. 


In the future, the SPEChpc~2021 suites can further be extended with applications from a broader range of domains. 
Due to the growing importance of application scaling and of scalable benchmarking in HPC, weak scaling benchmarks are needed that will prove to be beneficial for the HPC community, and the SPEChpc~2021 benchmarks could be extended in this direction. 
